# The Relationship between Creativity, Imitation, and Cultural Diversity

Liane Gabora and Stefan Leijnen
University of British Columbia

**Abstract**
There are both benefits and drawbacks to cultural diversity. It can lead to friction and exacerbate differences. However, as with biological diversity, cultural diversity is valuable in times of upheaval; if a previously effective solution no longer works, it is good to have alternatives available. What factors give rise to cultural diversity? This paper describes a preliminary investigation of this question using a computational model of cultural evolution. The model is composed of neural network based agents that evolve fitter ideas for actions by (1) inventing new ideas through modification of existing ones, and (2) imitating neighbors' ideas. Numerical simulations indicate that the diversity of ideas in a population is positively correlated with both the proportion of creators to imitators in the population, and the rate at which creators create. This is the case for both minimum and peak diversity of actions over the duration of a run.

## 1 Introduction

Computer science is drawing ever more extensively upon the natural world for inspiration in the design of search algorithms, optimization tools, problem solving techniques, and even computer-based artistic expression. What Mother Nature—probably the most effective problem solver—has come up with is the human mind itself. The brain's effectiveness derives largely from the fact that it is endlessly creative, able to break out of ruts and come up with ideas and solutions that are new, useful, and appealing. Not only are we individually creative, but we build on each other's creations such that over the centuries our ideas and inventions can be said to have evolved. In order for computer scientists to put to use the process by which creative ideas evolve through cultural exchange we must first develop better computational representations of the process. This paper investigates one of its aspects: the interaction between how creative individuals are, and how numerous they are in a society.

Our capacity for self-expression, for finding practical solutions to problems of survival, and coming up with aesthetically pleasing objects that delight the senses, all stem from the creative power of the human mind. However, there are also considerable drawbacks to creativity. A creative solution to one problem often generates other problems or unexpected negative side effects that may only become apparent after much has been invested in the creative solution. Moreover, creative individuals are more emotionally unstable and prone to affective disorders such as depression and bipolar disorder, and have a higher incidence of schizophrenic tendencies, than other segments of the population [3,9,27,28,41]. They are also more prone to abuse drugs and alcohol [19,20,34,36,37,38] as well as suicide [21]. Also, creative people often feel disconnected from others because they defy the crowd [40,42].

However, in a group of interacting individuals only a fraction of them need be creative for the benefits of creativity to be felt throughout the group. The rest can reap the benefits of the creator's ideas without having to withstand the dark aspects of creativity by simply copying, using, or admiring them. After all, few of us know how to build a computer, or write a symphony, or a novel, but they are nonetheless ours to use and enjoy when we please. In a society of interacting individuals capable of imitation, some members can capitalize on the benefits of creativity without incurring the drawbacks by merely imitating their creative peers. This opens up some interesting questions such as the following: How does the ratio of creators to imitators affect the transmission and resulting variety of ideas in the population? As globalization increases and migration creates communities of ever greater diversity, it becomes increasingly important to know the underlying dynamics of how these parameters interact. A diversity of actions and approaches is commonly associated with strife, turmoil, and unrest, but also with excitement and creativity.

Understanding cultural diversity is particularly relevant in times of rapid change when old approaches are at risk of becoming outdated. We have investigated this using an agent-based modeling approach. The agents are too rudimentary to suffer any of these affective penalties of creativity. The model does incorporate two negative consequences of creativity that are much more straightforward and no less relevant than affective disorders. First, innovation is a slow and frustrating process. Only few innovators are lucky enough to progress quickly, the rest are struggling and straggling behind. Unless, of course, they too get a chance to imitate what their best peers have achieved and build on that. In other words, it is not only the imitators who can benefit from copying; innovators do too.

Similarly, the model is too simple to incorporate the potential friction, competition, and clashes as well as, on the other hand, the kind of cultural vibrance and innovation that can be associated with high diversity [7,8,43]. Nor does it incorporate the kind of changing environment that would make cultural diversity of particular benefit. The goal at this stage is simply to gain an understanding of how diversity is affected by different ratios of creative to uncreative individuals, and different degrees of creativity within creative individuals.

Whereas in the earliest versions of the computer model used here, all agents were equally capable of both inventing and imitating [11], in a subsequent version greater individualization was possible [15,33]. Each agent could be a pure imitator, a pure creator, or something in between. We found that for low probabilities of invention for creators, the mean fitness of ideas increased as a function of the percentage of creators in the society, but for higher invention probabilities, the optimal ratio of creators to imitators followed an non-linear decreasing function. Thus as a general rule, the more creative the creators were, the less numerous they should be. In this paper we report on a more extensive investigation that employs a more detailed and analysis of these questions, and focuses more squarely on.

## 2 The Modeling Approach

EVOC consists of neural network based agents that invent ideas for actions, and imitate neighbors' actions [14,15,16]. EVOC is an elaboration of Meme and Variations, or MAV [11], the earliest computer program to model culture as an evolutionary process in its own right. MAV was inspired by the genetic algorithm (GA), a search technique that finds solutions to complex problems by generating a 'population' of candidate solutions through processes akin to mutation and recombination, selecting the best, and repeating until a satisfactory solution is found. Although MAV has inspired the incorporation of cultural phenomena (such as imitation, knowledge-based operators, and mental simulation) into evolutionary search algorithms [32], the goal behind MAV was not to solve search problems, but to gain insight into how ideas evolve. It used neural network based agents that could (1) invent new ideas by modifying previously learned ones, (2) evaluate ideas, (3) implement ideas as actions, and (4) imitate ideas implemented by neighbors. Agents evolved in a cultural sense, by generating and sharing ideas for actions, but not in a biological sense; they neither died nor had offspring. The approach can thus be contrasted with computer models of the interaction between biological evolution and individual learning [4,5,22,24,26].

MAV successfully modeled how descent with modification can occur in a cultural context, but it had limitations arising from the outdated methods used to program it. Moreover, although new ideas in MAV were generated making use of acquired knowledge and pattern detection, the name 'Meme and Variations' implied acceptance of the notion that cultural novelty is generated randomly, and that culture evolves through a Darwinian process operating on discrete units of culture, or 'memes'. Problems with memetics and other Darwinian approaches to culture have become increasingly apparent [6,10,12,15,16,29]. One problem is that natural selection prohibits the passing on of acquired traits (thus you don't inherit your mother's tattoo) [1]. In culture, however, 'acquired' change-that is, modification to ideas between the time they are learned and the time they are expressed-is unavoidable. Darwinian approaches must assume that elements of culture are expressed in the same form as that in which they are acquired. Natural selection also assumes that lineages do not intermix. However, because ideas cohabit a distributed memory with a multitude of other ideas, they are constantly combining to give new ideas, and their meanings, associations, and implications are constantly revised. It has been proposed that what evolves through culture is not discrete memes or artifacts, but the internal models of the world that give rise to them, and they evolve not through a Darwinian process of competitive exclusion but a Lamarckian process involving exchange of innovation protocols [13,14,15,16]. EVOC incorporates this in part by allowing agents to have multiple interacting needs, thereby fostering complex actions that fulfill multiple needs. Elsewhere [13,14] results of experiments using different needs and/or multiple needs are described, as well as how cultural evolution is affected by affordances of the agents' world, such as world shape and size, population density, and barriers that impede information flow, and potentially erode with time. This paper investigates how different proportions of creative to uncreative agents affects the fitness and diversity of ideas.

---

[1] That isn't to say that inheritance of acquired traits never occurs in biological evolution; it does. However to the extent that this is the case natural selection cannot provide an accurate model of biological evolution. Because inheritance of acquired traits is the exception in biology not the rule, natural selection still provides a roughly accurate model of biological evolution.

## 3 Architecture

EVOC consists of an artificial society of agents in a two-dimensional grid-cell world. The world used for these simulations is a toroidal lattice (or graph) with 100 nodes. Each node is occupied by a single, stationary agent. We used a von Neumann neighborhood structure, i.e. each agent interacts only with its four adjacent neighbors. We now describe the key components of the agents and the world they inhabit.

### 3.1 The agent

Agents consist of (1) a neural network, which encodes ideas for actions and detects trends in what constitutes a fit action, and (2) a body, which implements actions. In MAV there was only one need-to attract a mate. Thus actions were limited to gestures that attract mates. In EVOC agents can also engage in tool-making actions.

#### 3.1.1 The neural network

The core of an agent is a neural network, as shown in Fig. 1. It is composed of six input nodes that represent concepts of body parts (LEFT ARM, RIGHT ARM, LEFT LEG, RIGHT LEG, HEAD, and HIPS), six matching output nodes, and six hidden nodes that represent more abstract concepts (LEFT, RIGHT, ARM, LEG, SYMMETRY and MOVEMENT). Input nodes and output nodes are connected to hidden nodes of which they are instances (e.g., RIGHT ARM is connected to RIGHT.) Activation of any input node increases activation of the MOVEMENT hidden node. Opposite-direction activation of pairs of limb nodes (e.g., leftward motion of one arm and rightward motion of the other) activates the SYMMETRY node. The neural network learns ideas for actions. An idea is a pattern of activation across the output nodes consisting of six elements that instruct the placement of the six body parts. Training of the neural network is as per [11]. In brief, the neural network starts with small random weights, and patterns that represent ideas for actions are presented to the network. Each time a pattern is presented, the network's actual output is compared to the desired output. An error term is computed, which is used to modify the pattern of connectivity in the network such that its responses become more correct. Since the neural network is an autoassociator, training continues until the output is identical to the input. At this point training stops and the run begins. The value of using a neural network is simply that trends about what makes for a fit action can be detected using the symmetry and movement nodes (see below). The neural network can also be turned off to compare results to those obtained using instead of a neural network a simple data structure that cannot detect trends, and thus invents ideas at random.

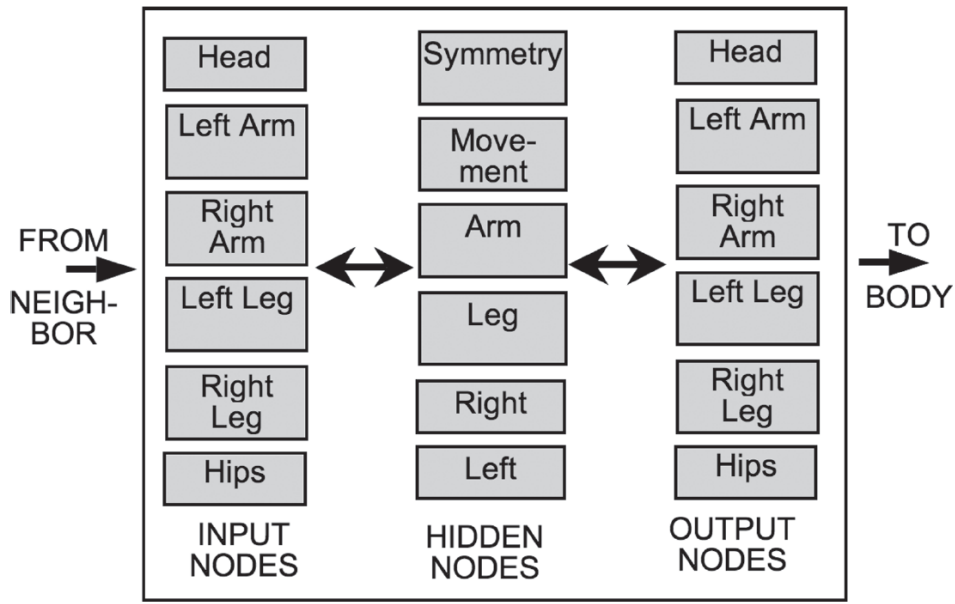

Figure 1. The neural network. See text for details.

**3.1.2 Knowledge-based operators**
Brains detect regularity and build schemas with which they adapt the mental equivalents of mutation and recombination to tailor actions to the situation at hand. Thus they generate novelty strategically, on the basis of past experience. Knowledge-based operators are a crude attempt to incorporate this into the model. Since a new idea for an action is not learned unless it is fitter than the currently implemented action, newly learned actions provide valuable information about what constitutes an effective idea. This information is used by knowledge-based operators to probabilistically bias invention such that new ideas are generated strategically as opposed to randomly. Thus the idea is to translate knowledge acquired during evaluation of an action into educated guesses about what makes for a Fit action. Two rules of thumb are used. The first rule is: if movement is generally beneficial, the probability increases that new actions involve movement of more body parts.

Each body part starts out at a stationary rest position, and with an equal probability of changing to movement in one direction or the other. If the fitter action codes for more movement, increase the probability of movement of each body part. Do the opposite if the fitter action codes for less movement. This rule of thumb is based on the assumption that movement in general (regardless of which particular body part is moving) can be beneficial or detrimental. This seems like a useful generalization since movement of any body part uses energy and increases the likelihood of being detected. It is implemented as follows:

```
am1 = movement node activation for current action
am2 = movement node activation for new action
p(im)i = probability of increased movement at body part i
p(dm)i = probability of decreased movement at body part i
IF (am2 > am1)
THEN p(im)i = MAX(1.0, p(im)i + 0.1)
```

```
ELSE IF (am2 < am1)
THEN p(im)i = MIN(0.0, p(im)i - 0.1)
p(dm)i = 1 - p(im)i
```

The second rule of thumb is: if fit actions tend to be symmetrical (e.g., left arm moves to the right and right arm moves to the left), the probability increases that new actions are symmetrical. This generalization is biologically sensible, since many useful actions (e.g., walking) entail movement of limbs in opposite directions, while others (e.g., pushing) entail movement of limbs in the same direction. This rule is implemented in a manner analogous to that of the first rule. In summary, each action is associated with a measure of its effectiveness, and generalizations about what seems to work and what does not are translated into guidelines that specify the behavior of the algorithm.

### 3.1.3 The body
If the fitness of an action is evaluated to be higher than that of any action learned thus far, it is copied from the output nodes of the neural network that represent concepts of body parts to a six digit array that contains representations of actual body parts, referred to as the body. Since it is useful to know how many agents are doing essentially the same thing, when node activations are translated into limb movement they are thresholded such that there are only three possibilities for each limb: stationary, left, or right. Six limbs with three possible positions each gives a total of 729 possible actions. Only the action that is currently implemented by an agent's body can be observed and imitated by other agents.

### 3.2 The fitness functions
Agents evaluate the effectiveness of their actions according to how well they satisfy needs using a pre-defined equation referred to as a fitness function. Agents have two possible needs. The fitness of an action with respect to the need to attract mates is referred to as F1, and it is calculated as in Ref. [11]. F1 rewards actions that make use of trends detected by the symmetry and movement hidden nodes and used by knowledge-based operators to bias the generation of new ideas. F1 generates actions that are relatively realistic mating displays, and exhibits a cultural analog of epistasis. In biological epistasis, the fitness conferred by the allele at one gene depends on which allele is present at another gene. In this cognitive context, epistasis is present when the fitness contributed by movement of one limb depends on what other limbs are doing. In these simulations F1 is used exclusively.

### 3.3 Incorporation of cultural phenomena
In addition to knowledge-based operators, discussed previously, agents incorporate the following phenomena characteristic of cultural evolution as parameters that can be turned off or on (in some cases to varying degrees):
- *Imitation.* Ideas for how to perform actions spread when agents copy neighbors' actions. This enables them to share effective, or 'fit', actions.
- *Invention.* This code enables agents to generate new actions by modifying their initial action or a previously invented or imitated action using knowledge-based operators (discussed previously).
- *Mental simulation.* Before implementing an idea as an action, agents can use the

fitness function to assess how fit the action would be if it were implemented.

**3.4 The world**
MAV allowed only worlds that were toroidal, or wrap-around. Moreover, the world was always maximally densely populated, with one stationary agent per cell. In EVOC the world can be either toroidal or square, and as sparsely or densely populated as desired, with stationary agents placed in any configuration. EVOC also allows the creation of complete or semi-permeable borders, which may be permanent or eroding. (This limits the probability of agents imitating others from different enclaves.)

**3.5 A typical run**
Each iteration, every agent has the opportunity to (1) acquire an idea for a new action, either by imitation, copying a neighbor, or by invention, creating one anew, (2) update the knowledge-based operators, and (3) implement a new action. To invent a new idea, the current action is copied to the input layer of the neural network, and this previous action is used as a basis from which to generate a new one. For each node the agent makes a probabilistic decision as to whether change will take place. If it does, the direction of change is stochastically biased by the knowledge- based operators using the activations of the SYMMETRY and MOVEMENT nodes.

Mental simulation is used to determine whether the new idea has a higher fitness than the current action. If so, the agent learns and implements the action specified by the new idea. To acquire an idea through imitation, an agent randomly chooses one of its neighbors, and evaluates the fitness of the action the neighbor is implementing using mental simulation. If its own action is fitter than that of the neighbor, it chooses another neighbor, until it has either observed all of its immediate neighbors, or found one with a fitter action. If no fitter action is found, the agent does nothing. Otherwise, the neighbor's action is copied to the input layer, learned, and implemented. Fitness of actions starts out low because initially all agents are immobile. Soon some agent invents an action that has a higher fitness than doing nothing, and this action gets imitated, so fitness increases. Fitness increases further as other ideas get invented, assessed, implemented as actions, and spread through imitation. The diversity of actions initially increases due to the proliferation of new ideas, and then decreases as agents hone in on the fittest actions.

**3.6 Implementation**
EVOC is written in Java, an object oriented programming environment, using the Joone open source neural network library. The graphical user interface makes use of the open-source charting project, JFreeChart, enabling variables to be user defined at run time, and results to become visible as the computer program runs.

**4 Summary of Previous Results**
EVOC closely replicates the results of experiments conducted with MAV [11]. The graph on the bottom left of Fig. 2 shows the increase in fitness of actions. The graph on the bottom right of Fig. 2 shows the increase and then decrease in the diversity of actions. Other results include:

- Fitness increases most quickly with an invention to imitation ratio of about 2:1.

- For the agent with the fittest actions, however, the less it imitates, the better it does.
- Increasing the invention-to-imitation ratio increases the diversity of actions. If increased much beyond 2:1, it takes more than twice as many iterations for all agents to settle on optimal actions.
- As in biology, epistatically linked elements take longer to optimize. (As explained earlier, in the present context epistasis refers to the situation where the effect on fitness of what one limb is doing depends on what another is doing.)
- The program exhibits drift-the biological term for change in the relative frequencies of alleles (forms of a gene) as a statistical byproduct of randomly sampling from a finite population [44]. With respect to culture, it pertains to possible forms of a component of an idea (e.g. if the idea is to implement the gesture 'wave', this can be executed with the left or right hand).
- Diversity of actions is positively correlated with number of needs that agents attempt to satisfy.
- Diversity of actions is positively correlated with population size and density, and with barriers between populations.
- Square (as opposed to toroidal) worlds foster higher diversity, as idea propagation is impeded by corners and edges.
- Slowly eroding borders increase fitness without sacrificing diversity by fostering specialization followed by sharing of fit actions.
- Introducing a leader that broadcasts its actions throughout the population increases the fitness of actions but reduces diversity of actions. Increasing the number of leaders reduces this effect.

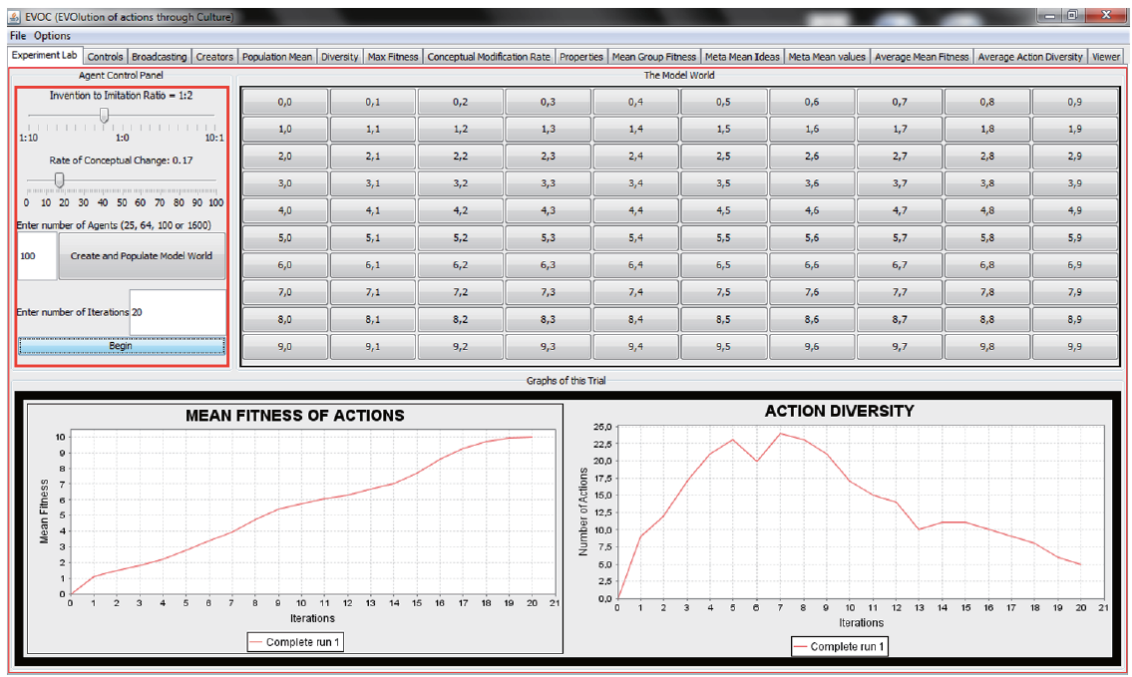

Figure 2. Output panel of GUI using F1. See text for details.

## 5 Experiments

In previous experiments [11,15,16], all agents have an equal probability of inventing or imitating. The choice of action is determined by a number of different factors, and the role an agent may take can vary each time step. Here we make a distinction between two types of agents, which differ in the extent to which they are creative, i.e., able to invent new actions by modifying previous ones. Whereas one kind of agent, referred to as imitators, only obtains new actions by imitating neighbors, the other type of agent, referred to as inventors or creators, will imitate neighbors at some times but generally obtain new actions by inventing them. Imitation only occurs when it leads to an improvement of the current action. All new actions in these simulations are generated by the creative agents; that is, imitators do not invent at all. Imitators simply copy the successful innovations of the creative agents, and thereby serve as a 'memory' for preserving the fittest configurations.

We investigated both the effect of varying how abundant creators are, and the effect of varying how creative they are. The frequency of creators (that is, the proportion of creators relative to imitators in the population--is referred to as C.

The creativity of creators (that is, the probability that a creator invents a new action instead of imitating a neighbor) is referred to as p. It may help to think of the entire population as being divided in three subgroups at any given iteration:

- $C * p * 100$ agents are creative agents attempting to innovate;
- $C * (1 – p) * 100$ agents are creative but not attempting to innovate;
- $(1 – C) * 100$ agents always imitating.

It is also important to keep in mind that whereas the attributes creative and imitative are permanent, the subgroups of creative agents either innovating or imitating at any given time fluctuate stochastically. The process of innovation is driven by the neural network explained previously. If the attempt to innovate is abortive, the agent retains it current configuration and fitness. If a creative agent does not attempt to innovate (with a probability of $1 – p$), it will behave as an imitator. The process of imitation is analogous to lazy (non-greedy) search." The imitating agent will scan its four neighbors in random order, adopt the first configuration with a fitness greater than its own or, failing so, retain its current status.

In these simulations there are two negative consequences of creativity. The first is that an iteration spent inventing is an iteration not spent imitating. The second is that creative change can break up co-adapted partial solutions. Actions have a cultural version of what in biology is referred to as epistasis, wherein what is optimal with respect to one part depends on what is done with respect to another part. Once both parts of the problem have been solved in a mutually beneficial way, too much creativity can cause these co-adapted solutions to break down.

In the simulations reported on here, the world is not segmented, i.e. there are no barriers as in several other EVOC experiments reported elsewhere. Broadcasting is not used, and the probability of change in any single body part is 1/62. Given the six body parts the choice of 1/6 as the probability of change in any one of them maximizes the probability that one and only one body part will change at a time [16]. All experiments feature runs of 100 iterations and results displayed are averaged over 100 runs; on each run, the creative agents are randomly dispersed.

Figure 3 illustrates the average diversity, or number of different actions in the artificial society, over the course of a run as a function of C and p. Action diversity is positively correlated with both the percentage of creators, and their level of creativity. As C and p increase, a larger fraction of the search space is discovered. This generally holds true for both the maximum (peak) diversity as well as the minimal diversity during a run.

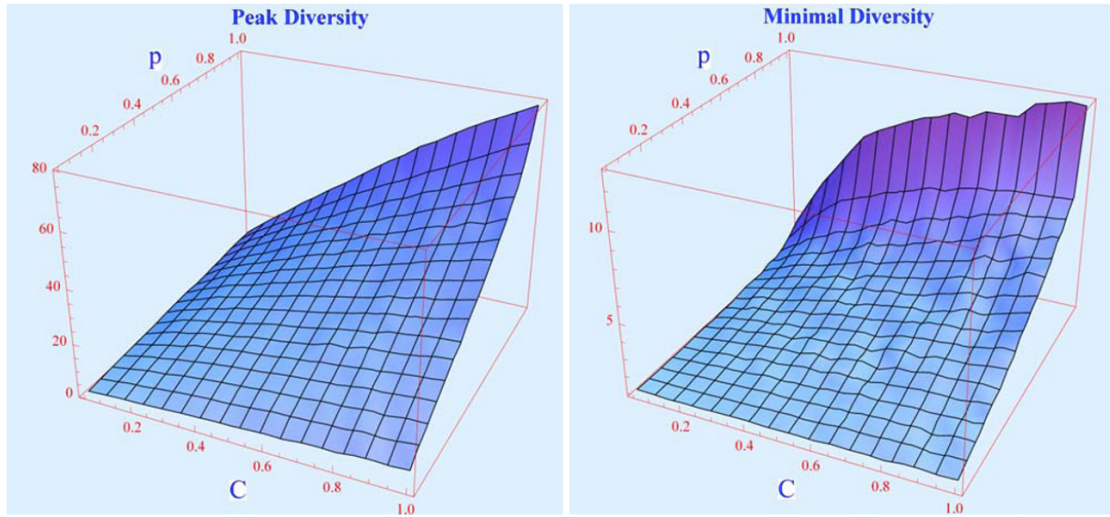

Figure 3. Average action diversity for different values of creator innovation probability (p) and creator-to-imitator ratio (C), showing maximum (left) and minimum (right) diversity. High levels of diversity above eight (the number of optimal actions) are negatively correlated with population fitness. Also, the absence of recombination lowers the potential impact of a high idea diversity. (In future implementations, when we explore the impact of a changing fitness landscape and recombination, it is expected that it will no longer necessarily be the case that high diversity is necessarily correlated with low fitness.).

**Discussion**
The goal of the work reported here was to investigate how creativity and imitation interact to generate a pattern of ideas across a society, focusing on the diversity of ideas. The experiments reported here incorporate two potentially negative consequences of creativity. The first is that an iteration spent inventing is an iteration not spent imitating. The second is that creative change can break up co-adapted partial solutions.

In earlier experiments conducted with EVOC, the neural network based computer model of cultural evolution used here, all agents were equally capable of both inventing and imitating [11]. In subsequent experiments the code was modified to allow some agents to be creators and others imitators [33]. This enabled us to investigate a related but different question. Since in a society of interacting individuals capable of imitation, some members can capitalize on the benefits of creativity without incurring the drawbacks by merely imitating their creative peers, what is the ideal ratio of creators to imitators?

In this paper we found action diversity to be positively correlated with both the percentage of creators, C, and their level of creativity, p. This was the case both for the peak diversity obtained at any point throughout a run, as well as the number of actions the artificial society eventually settled upon, which we refer to as the minimal diversity. in future studies we plan to compare these findings to real world data. The findings have important implications in the context of the current rapidly changing cultural environment. As in biological evolution, diversity is of little value in times of stasis, when it is best for the population as a whole to simply converge on optimal solutions. However, diversity is very important in times of change, where the optimal solution today can be ineffective tomorrow, and it is valuable to have a variety of alternative solutions on hand.

We note that the results obtained here reflect in part limitations in the architecture of the model. EVOC in its current implementation does not accommodate combinatorial or selective imitation. An agent either copies exactly what another agent or doing or ignores that other agent; it cannot choose bits and pieces that would augment or complement elements of its current action. Nor can an agent copy selectively combining elements of the actions exhibited by agents at once. Consequently, imitation while it is essential to the rapid spread for superior configurations destroys diversity. We expect that combinatorial or selective copying will temper the destructive effects of imitation. We will study also investigate how it affects the fitness or effectiveness of ideas across the population.

Indeed a major topic for future research is to look at the effect of changing the ratio of creators to imitators, and the degree to which creators are creative, on not just cultural diversity but the fitness of ideas. Particularly interesting in this regard will be to investigate how diversity and fitness of ideas responds to a changing fitness landscape.

**Acknowledgments**
This research is supported by grants to Liane Gabora from the Natural Sciences and Engineering Research Council of Canada and the Scientific Research Fun of the Flemish Government of Belgium. We thank Tiha von Ghyczy for assistance with the analysis.